\documentclass[a4paper, 10pt, conference]{ieeeconf}      
\usepackage{FG2023}

\usepackage{times}
\usepackage{epsfig}
\usepackage{graphicx}
\usepackage{amsmath}
\usepackage{amssymb}
\usepackage{booktabs}
\usepackage{tabularx}
\usepackage{multirow,bigdelim}
\usepackage{hyperref}

\newcommand{\com}[1]{}

\newcommand{\figref}[1]{Fig.~\ref{fig:#1}}
\newcommand{\tblref}[1]{Table~\ref{tbl:#1}}
\newcommand{\secref}[1]{Section~\ref{sec:#1}}

 \FGfinalcopy 

\IEEEoverridecommandlockouts                              
\def\FGPaperID{2} 

\title{\LARGE \bf Automatic Assessment of Infant Face and Upper-Body Symmetry \\ 
as Early Signs of Torticollis}

\vspace{-2mm}

\author{\parbox{16cm}{\centering
    {\large Michael Wan$^{1,2}$, Xiaofei Huang$^2$, Bethany Tunik$^3$, Sarah Ostadabbas$^{2,4}$}\\
    {\normalsize
    $^1$Roux Institute, Northeastern University, Portland, ME, USA\\
    $^2$Augmented Cognition Lab (ACLab), Northeastern University, Boston, MA, USA\\
    $^3$Board-Certified Clinical Specialist in Pediatric Physical Therapy, Boston, MA, USA\\
    $^4$Corresponding author: \texttt{ostadabbas@ece.neu.edu}}}
}

\vspace{-2mm}

\newcommand{\tblgeometricquantities}{
\begin{table*}[ht]
\caption{Geometric measures of face and upper body symmetry pertaining to torticollis (details in \secref{concepts})} 
\vspace{-0.2in}
\begin{center}
        \begin{tabular}{llp{2.8in}ll}
        \toprule
        \textbf{Measure} & \textbf{Code} & \textbf{Description} & \textbf{Definition} & \textbf{Source}\\ 
        \midrule
        Orbit slopes angle & osa & Angle between outer cathnus (eye corner) line and inner cathnus line & $\measuredangle\left(\overrightarrow{P_{36}P_{45}}, \overrightarrow{P_{39}P_{42}}\right)$ & \cite{akbari_facial_2015,HELVESTON19921609}\\
        \midrule
        Relative face size & rfs & Left divided by right outer-canthus-to-mouth-corner lengths & $\left\Vert\overrightarrow{P_{45}P_{54}}\right\Vert\bigg/\left\Vert\overrightarrow{P_{36}P_{48}}\right\Vert$ & \cite{akbari_facial_2015,HELVESTON19921609}\\
        \midrule
        Facial angle & fa & Angle between the eye line and mouth corners line & $\measuredangle\left(\overrightarrow{P^\mu_{36,39}P^\mu_{42,45}}, \overrightarrow{P_{48}P_{54}}\right)$ & \cite{akbari_facial_2015,wilson_hoxie_1993}\\
        \midrule
        Gaze angle & ga & Angle between outer cathnus line and midsternal plumb line & $\measuredangle\left(\overrightarrow{P_{36}P_{45}}, {\perp}\overrightarrow{P_{68}P_{69}}\right)$ & \cite{bhaskar_congenital_2017} \\
        \midrule
        Translational deformity & td & Distance between chin and midsternal plumb line, normalized by the distance between the outer cathnuses & $\left\Vert P_8,{\perp}\overrightarrow{P_{68}P_{69}}\right\Vert\bigg/\left\Vert\overrightarrow{P_{36}P_{45}}\right\Vert$ & \cite{bhaskar_congenital_2017}\\
        \midrule
        Habitual head deviation & hhd & Angle between eye line and acromion process (shoulder) line & $\measuredangle\left(\overrightarrow{P^\mu_{36,39}P^\mu_{42,45}}, \overrightarrow{P_{68}P_{69}}\right)$ & \cite{rahlin_tamo_2005,rahlin_reliability_2010}\\
        \bottomrule
        \end{tabular}
\label{tbl:geometric-quantities}
\vspace{-0.2in}
\end{center}
\end{table*}
}

\newcommand{\tblestimators}{
\begin{table}
\caption{Configuration of our landmark estimation models}
\vspace{-0.3in}
\begin{center}
        \begin{tabular}{llll}
        \toprule
        Model & Task  & Backbone & Training Data\\ 
        \midrule
        HRNet \cite{WangSCJDZLMTWLX19} & Adult Face & HRNet & 300-W Train \cite{SAGONAS20163}\\
        DarkPose \cite{Zhang_2020_CVPR} & Adult Body & HRNet & COCO \cite{Lin_2014_COCO}\\
        HRNet-R90JT \cite{wan_infanface_2022} & Infant Face & HRNet & InfAnFace Train \cite{wan_infanface_2022}\\
        FiDIP \cite{huang_invariant_2021} & Infant Body & DarkPose & SyRIP \cite{huang_invariant_2021}\\
        \bottomrule
        \end{tabular}
\label{tbl:estimators}
\vspace{-0.3in}
\end{center}
\end{table}
}

\newcommand{\tblperformance}{
\begin{table}[t]
\caption{Estimation performance of geometric measures of symmetry from infant and adult-based models} 
\vspace{-0.3in}
\begin{center}
\resizebox{\columnwidth}{!}{
\begin{tabular}{lllrrrrrr}
\toprule
\textbf{Metric} & & \textbf{Models} & \textbf{osa} & \textbf{rfs} & \textbf{fa} & \textbf{ga} & \textbf{td} & \textbf{hhd}\\
\midrule
\multirow{4}{*}{Spearman's $\rho$} & \multirow{4}{*}{$\uparrow$} & \multirow{2}{*}{Infant} & \textbf{0.36} & \textbf{0.61} & \textbf{0.60} & \textbf{0.79} & \textbf{0.53} & \textbf{0.80}\\
&&& $\ast$ & $\ast{\ast}\ast$ & $\ast{\ast}\ast$ & $\ast{\ast}\ast$ & $\ast{\ast}\ast$ & $\ast{\ast}\ast$\\
& & \multirow{2}{*}{Adult} & -0.10 & 0.58 & 0.53 & 0.78 & 0.36 & 0.78\\ 
&&&& $\ast{\ast}\ast$ & $\ast{\ast}\ast$ & $\ast{\ast}\ast$ & $\ast$ & $\ast{\ast}\ast$\\
\midrule
\multirow{2}{*}{BCA ($\%$)} & \multirow{2}{*}{$\uparrow$} & Infant & \textbf{50.0} & \textbf{77.8} & \textbf{75.0} & \textbf{80.6} & \textbf{88.9} & \textbf{77.8}\\
 & & Adult & 41.7 & 75.0 & 66.7 & 77.8 & 69.4 & \textbf{77.8}\\
\midrule
\multirow{2}{*}{MAE} & \multirow{2}{*}{$\downarrow$} &  Infant & \textbf{1.98} & \textbf{0.04} & \textbf{2.18} & \textbf{2.29} & \textbf{0.06} & \textbf{2.63}\\
& & Adult & 4.73 & 0.07 & 4.51 & 5.25 & 0.08 & 4.71\\
\midrule
\multirow{2}{*}{RMSE} & \multirow{2}{*}{$\downarrow$} & Infant & \textbf{2.70} & \textbf{0.05} & \textbf{3.38} & \textbf{3.28} & \textbf{0.09} & \textbf{3.51}\\
& & Adult & 12.88 & 0.17 & 11.27 & 15.61 & 0.12 & 11.71\\
\midrule
\multicolumn{3}{l}{Measure type} & $^{\circ}$ & ratio & $^{\circ}$ & $^{\circ}$ & ratio & $^{\circ}$\\
\multicolumn{3}{l}{Ground truth mean} & 0.80 & 1.00 & -0.34 & 87.64 & 0.11 & -2.62\\
\midrule
\multicolumn{8}{l}{$\ast$ $p\leq 0.05$, ${\ast}{\ast}$ $p\leq0.01$, $\ast{\ast}\ast$ $p\leq0.001$; best performance in \textbf{bold}} \\
\bottomrule
\end{tabular}}
\label{tbl:performance}
\vspace{-0.3in}
\end{center}
\end{table}
}

\newcommand{\tblspearman}{
\begin{table}
\caption{Interpretation of Spearman's $\rho$, from \cite{dancey2007statistics}}
\vspace{-0.2in}
\begin{center}
        \begin{tabular}{ll}
        \toprule
        \textbf{Spearman's} $\rho$ & \textbf{Correlation}\\ 
        \midrule
        $\geq$ 0.70 & Very strong relationship\\
        0.40--0.69 & Strong relationship\\
        0.30--0.39 & Moderate relationship\\
        0.20--0.29 & Weak relationship\\
        0.01--0.19 & No or negligible relationship\\
        \midrule
        \multicolumn{2}{c}{Descriptions apply to both positive and negative values}\\
        \bottomrule
        \end{tabular}
\label{tbl:spearman}
\vspace{-0.3in}
\end{center}
\end{table}
}

\newcommand{\figgeometricquantities}{
\begin{figure}
 \centering
 \includegraphics[width=1.0\linewidth]{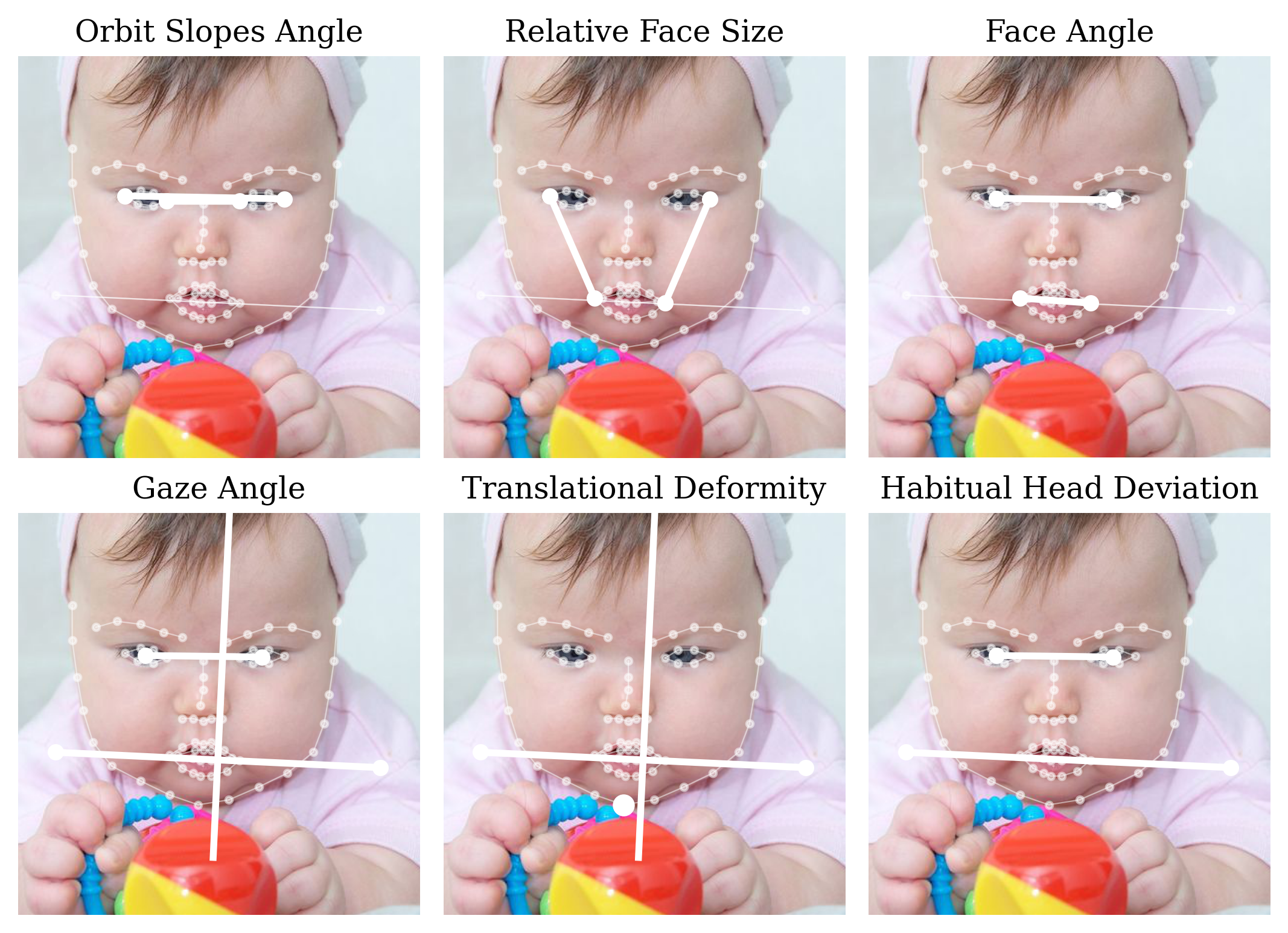}
 \vspace{-0.2in}
 \caption{We study the effectiveness of using deep learning computer vision techniques to assess a range of geometric facial and upper body measures of symmetry---illustrated schematically here---which are drawn from medical research literature on torticollis in infants and children. The assessments employ recent advances in infant-domain estimation of facial and upper body landmarks---also illustrated faintly---and we demonstrate that this yields better results than landmark estimation methods largely trained on adult data. }
  \vspace{-0.2in}
 \label{fig:geometric-quantities}
\end{figure}

}

\newcommand{\figfacenumbers}{
\begin{figure}
 \centering
 \includegraphics[width=1.0\linewidth]{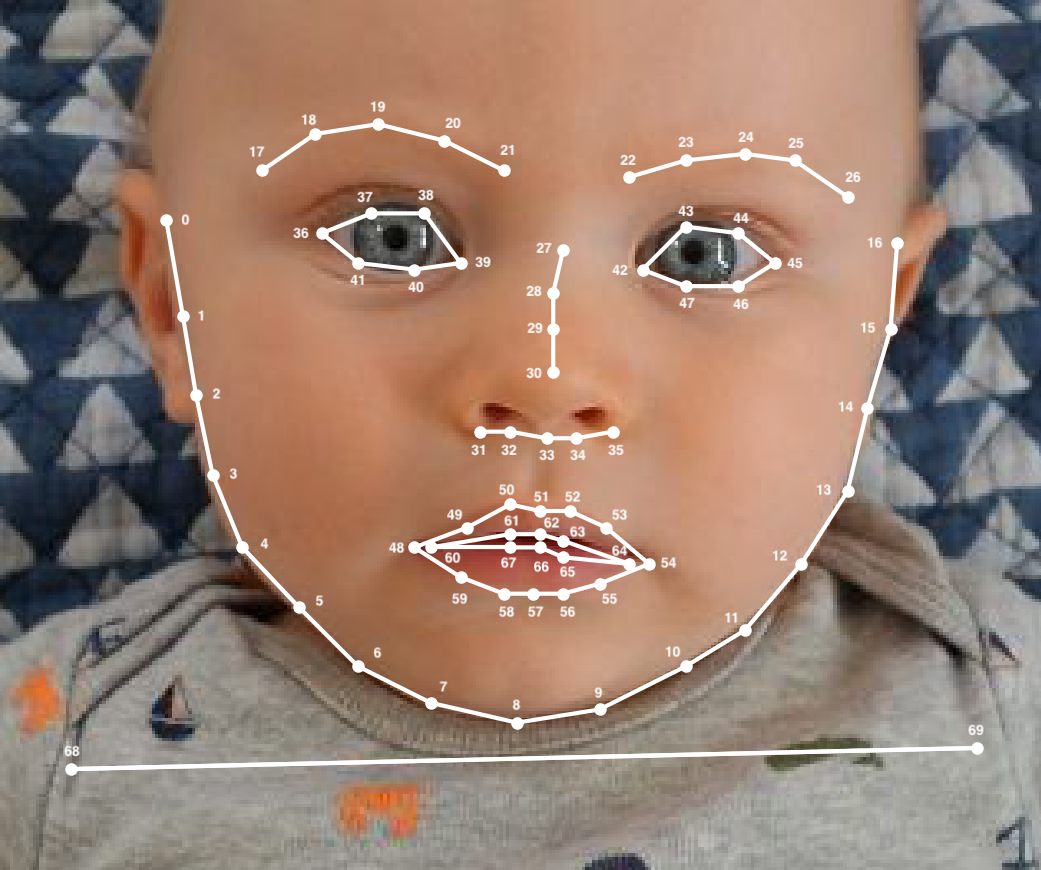}
 \vspace{-0.2in}
 \caption{Face and body (shoulder) landmarks available from ground truth annotations, with index numbers for each landmark point in correspondence with the definitions in \tblref{geometric-quantities}.}
 \vspace{-0.2in}
 \label{fig:face-numbers}
\end{figure}
}

\newcommand{\figscatter}{
\begin{figure*}[ht]
 \centering
 \includegraphics[width=0.9\linewidth]{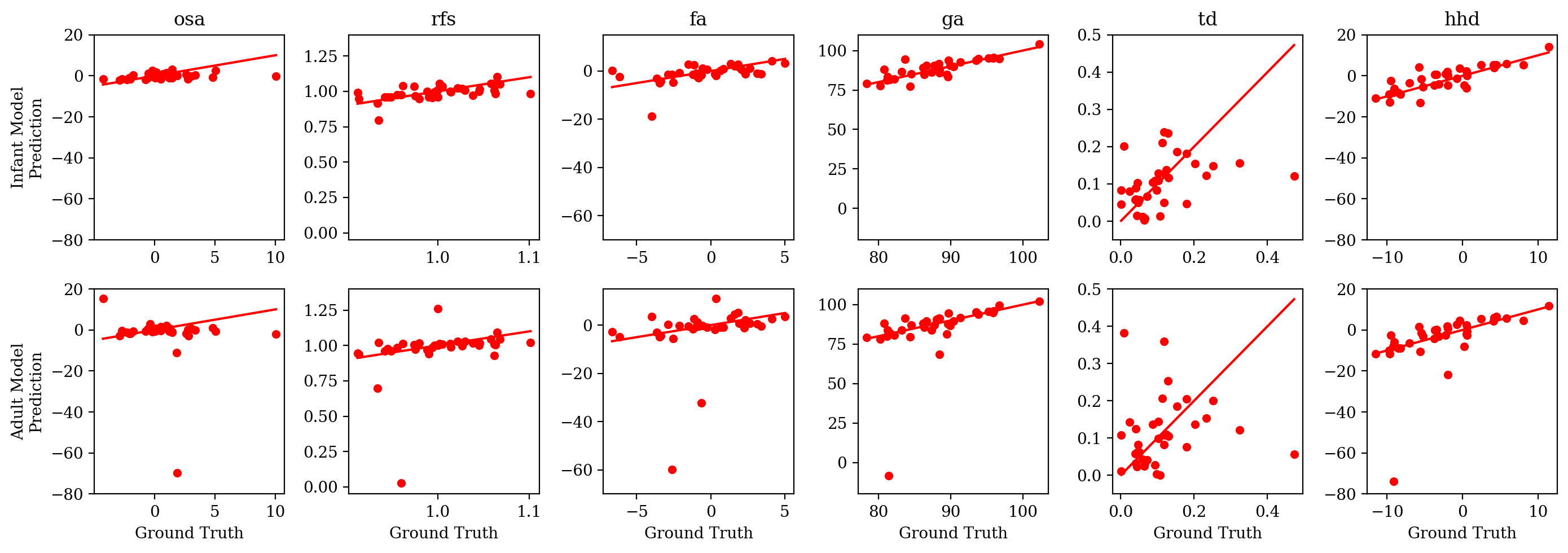}
 \vspace{-0.1in}
 \caption{Scatter plots of predictions vs. ground truth of our six geometric measures of symmetry (definitions given in \tblref{geometric-quantities}), with the scale chosen to emphasize the strong effect of outliers on the predictions derived from adult pose estimation models (bottom) compared to those derived from the infant pose estimation models (top). See \tblref{performance} for a more quantitative characterization of performance for each quantity. Identity lines drawn for reference.}
 \vspace{-0.1in}
 \label{fig:scatter}
\end{figure*}
}

\newcommand{\figpredictions}{
\begin{figure*}[ht]
 \centering
 \includegraphics[width=0.9\linewidth]{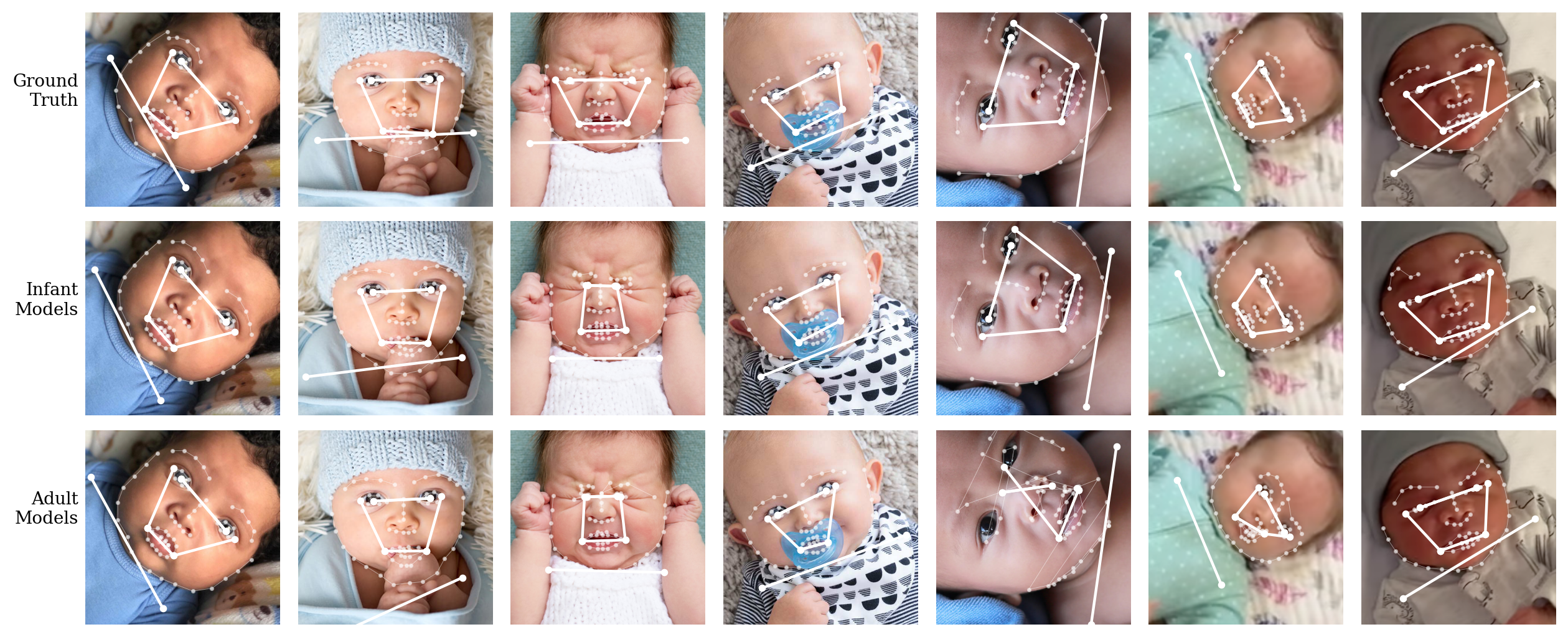}
 \vspace{-0.1in}
 \caption{Ground truth and predicted positions of some geometric elements related to our measures of symmetry, superimposed on the underlying face and shoulder landmark estimations. Samples chosen to illustrate outlier cases of bad facial landmark predictions, which disproportionally contribute to the poor quantitative performance from the pose estimation based models, especially for the models trained on adult data. Outside of these outlier cases, performance is difficult to adjudicate by visual inspection, and thus our results analysis in \secref{results} is guided by quantitative performance metrics.}
 \vspace{-0.2in}
 \label{fig:predictions}
\end{figure*}
}

\begin{document}

\ifFGfinal
\thispagestyle{empty}
\pagestyle{empty}
\else
\author{Anonymous FG2023 submission\\ Paper ID \FGPaperID \\}
\pagestyle{plain}
\fi
\maketitle

\begin{abstract}

We apply computer vision pose estimation techniques developed expressly for the data-scarce infant domain to the study of torticollis, a common condition in infants for which early identification and treatment is critical. Specifically, we use a combination of facial landmark and body joint estimation techniques designed for infants to estimate a range of geometric measures pertaining to face and upper body symmetry, drawn from an array of sources in the physical therapy and ophthalmology research literature in torticollis. We gauge performance with a range of metrics and show that the estimates of most these geometric measures are successful, yielding strong to very strong Spearman's $\rho$ correlation with ground truth values. Furthermore, we show that these estimates, derived from pose estimation neural networks designed for the infant domain, cleanly outperform estimates derived from more widely known networks designed for the adult domain\footnote{Code and data available at \href{https://github.com/ostadabbas/Infant-Upper-Body-Postural-Symmetry}{https://github.com/ostadabbas/Infant-Upper-Body-Postural-Symmetry}. 979-8-3503-4544-5/23/\$31.00 ©2023 IEEE}.
\end{abstract}

\section{INTRODUCTION}
\vspace{-1mm}

Torticollis is a common condition in infants and children, characterized by a persistent neck tilt or twist to one side. Its most common form, congenital muscular torticollis (CMT), has an estimated incidence of 3.9\% to 16\% \cite{kaplan_physical_2018}. Early treatment is critical: outcomes are best when CMT is diagnosed and physical therapy treatment started before the infant is three months old, and conversely, if untreated or treated later, CMT can lead to face, skull, or spine deformities, pain and limited motion, and the need for invasive interventions and surgery \cite{sargent_congenital_2019}. Screening, diagnosis, and monitoring during treatment for CMT require laborious professional assessments, so with the recent onset of computer vision technology specifically studying infant face and body poses for health and developmental applications \cite{chambers_computer_2020, huang_invariant_2021, huang_computer_2022, liu_simultaneously-collected_2020, wan_infanface_2022}, it is natural to wonder whether algorithmic techniques can help enable remote monitoring or automated screening and diagnosis to augment clinical expertise. In this paper, as a first step towards such applications, we explore viability of using computer vision techniques to assess a set of geometric measures of symmetry in the face and upper body, previously identified in the medical literature as being relevant to CMT or the similarly presenting (non-congenital) ocular torticollis condition, purely from casual photographs of infants in their natural environments. 

\figgeometricquantities

The geometric symmetry measures we consider are illustrated in \figref{geometric-quantities}. We carefully researched and selected these from among measures studied by physical therapy and ophthalmology researchers to enable qualitative assessment of changes over time and in response to treatments \cite{akbari_facial_2015,HELVESTON19921609,rahlin_tamo_2005,wilson_hoxie_1993}, including specifically to quantify outcomes after surgery \cite{bhaskar_congenital_2017}. In \cite{rahlin_reliability_2010}, the authors even study the reliability of the procedure of extracting such measurements from still photographs itself. The general contention is not, of course, that measurements from individual still photographs are fully determinative of torticollis conditions, but rather that aggregated over time they can be employed alongside other tools as part of its detection and treatment. We hope that by establishing the viability of algorithmic assessments of these geometric measurements of symmetry from still photographs, we will open the door to future applications in automated monitoring and screening, including from videos.

The technical tools that we employ for this proof-of-concept experiment are based in computer vision pose estimation from still images, and in particular, face and body landmark estimation. Mature solutions to these tasks exist but are generally based in deep learning from primarily adult faces and bodies. As alluded earlier, specialized methods tailored for the unique faces and bodies of infants have only begun to crop up in recent years, in recognition of the significant domain gap between infant and adults from the point of view of computer vision representations. In this paper, we employ an infant face landmark estimation model from \cite{wan_infanface_2022} together with a infant body landmark estimation model from \cite{huang_invariant_2021}, both of which employ domain adaptation techniques to tune existing adult-focused models to the infant domain. We work with a subset of the InfAnFace dataset from \cite{wan_infanface_2022}, and compare the values of our six geometric symmetry measures derived from both predicted and ground truth landmarks. We propose modifications of these measures from their original definitions in the literature to enable compatibility with the landmarks used in pose estimation techniques. 

Our findings show that predictions derived from the recent infant-domain pose models exhibit ``strong'' or ``very strong'' Spearman's $\rho$ ranked correlation with the ground truth values on a precisely labeled test set of infant faces in the wild, with best performance on the gaze angle (ga) between the line connecting the outer corners of the eyes and the midsternal plumb line, and on the habitual head deviation (hhd), the angle between the eye line and the acromion process (shoulder) line. Predictions of three other measurements (including non-angles) were strong, but we found only moderate success in the predictions of the orbit slopes angle (osa), the angle between the lines connecting the outer and inner corners of the eyes---this being arguably the most subtle metric. Based on this and our further analysis involving more performance metrics in \secref{results}, we conclude that computer vision infant pose estimation techniques can successfully measure a range of quantities pertaining to torticollis.

\section{BACKGROUND: TORTICOLLIS AND INFANT DEVELOPMENT}
\label{sec:background}
\vspace{-1mm}


\subsection{Quantifying torticollis}
\label{sec:background-quantification}
\vspace{-1mm}

While there is a large corpus of research and established methodology in the diagnosis and treatment of torticollis, it is largely based on in-person physical assessments by experts and follow-ups with imaging or other deeper techniques \cite{kaplan_physical_2018}. By contrast, our work is inspired by a smaller cluster of papers dealing explicitly with measuring signs and symptoms of torticollis geometrically from still images. 

We start with congenital muscular torticollis (CMT). The author in \cite{rahlin_tamo_2005} studied the effectiveness of a specific therapeutic intervention for CMT by comparing changes in an infant's ``habitual head deviation'' (also ``head tilt'') as measured by hand from still photographs. The same author later studied the reliability of this photograph-based method itself, in \cite{rahlin_reliability_2010}. Separately, \cite{bhaskar_congenital_2017} measured the ``gaze angle'' and ``transformational deformity'' of child subjects, again from still photographs, to gauge improvement in response to surgical intervention. Measurements from photographs offer researchers a repeatable, objective way to quantify the change in severity of torticollis after an intervention. 

Sometimes torticollis is not congenital (present from birth) but rather acquired, as is the case with ocular torticollis, where the abnormal head posture is adopted to compensate for a defect in vision. In such cases, diagnoses often occur only in adulthood, and can be informed by examination of head pose in childhood photographs. Correspondingly, ophthalmologists have also studied the quantification of facial asymmetry via geometric measurements from still images. An overview of such methods and quantities is given in \cite{akbari_facial_2015}, who in turn cite \cite{HELVESTON19921609} for definitions of facial measurements such as the ``orbit slopes angle,'' ``relative face size,'' and ``facial bulk mass,'' and \cite{wilson_hoxie_1993} for definitions of the ``facial angle'' and the ``nasal tip deviation.'' These measurements are studied not as a means to quantify the effect of interventions, but as part of a more comprehensive study on the differential diagnosis of ocular torticollis and other conditions related to facial asymmetries, especially superior oblique palsy \cite{akbari_facial_2015}.

\subsection{Computer vision for infant health and development}
\vspace{-1mm}

We are not aware of prior computer vision research intended to detect torticollis or gauge head asymmetry in infants. The closest in spirit might be a pair of related papers, \cite{VU2022120154,zhang_hold_2022}, in which researchers employ computer vision to analyze head posture and tremor with a view towards algorithmic understanding of cervical dystonia (also known as spasmodic torticollis), with incidence largely in the adult population. Accordingly, those studies can take advantage of far more mature adult-focused head pose estimation techniques like OpenFace 2.0 \cite{Baltrusaitis2018OpenFace}, whereas our efforts are highly constrained by data scarcity in the infant domain. In the infant domain, there is prior work on bodies but not faces: \cite{chambers_computer_2020} develops an infant-specific body pose estimation deep network to extract body motion information from infant videos, in an attempt to assess infant neuromotor risk; and \cite{huang_computer_2022} uses 3D infant pose estimation techniques to assess infant body symmetry, with a view towards applications in infant development and torticollis, but without a specific implementation of such. As mentioned, all of this work comes in the context of recent attention in computer vision to the small data domain problem of infant pose comprehension, for both faces \cite{wan_infanface_2022} and bodies \cite{hesse_learning_2018,huang_invariant_2021,Huang2022Appearance,liu_simultaneously-collected_2020}.

\section{CONCEPTS: MEASUREMENTS OF SYMMETRY}
\label{sec:concepts}
\vspace{-1mm}

\tblgeometricquantities


\figfacenumbers

From the sources cited in \secref{background-quantification}, we identified all clearly defined geometric symmetry measures used by researchers in the study of torticollis and facial asymmetries---six in all. We altered the definitions to base them explicitly on the 68 facial landmark coordinates and two body joint (shoulder) coordinates used by our pose estimators, as illustrated and enumerated in \figref{face-numbers}, which also enables more consistent comparisons. The final measures are defined in \tblref{geometric-quantities} and illustrated in \figref{geometric-quantities}. In the rest of this section, we clarify and discuss these definitions.

\subsection{Assumptions and context}
\label{sec:assumptions}
\vspace{-1mm}
Since we work with both ground truth and estimated landmark coordinates in two dimensions, we generally assume that every infant is facing forward into the camera, so that the infant's face plane is roughly parallel with the camera image plane. In principle, all of the measurements we consider are well-defined for three dimensional face and bodies, but we do not have access to three dimensional face landmarks in the infant domain, and thus we must assume this alignment to ensure that these measurements are well-defined from the two dimensional landmarks. 

\subsection{Symbols and functions}
\vspace{-1mm}
In \tblref{geometric-quantities}, $\measuredangle(u,v)$ represents the signed angle in degrees between vectors $u$ and $v$, relative to a fixed orientation (say, clockwise)\footnote{We want to measure the \textit{signed} angle because, for instance, we would like a clockwise and a counterclockwise angles of equal magnitude to be considered different, for the purposes of quantifying predictions. Note that we have $\measuredangle(u,v) = - \measuredangle(v,u)$ for all $u$ and $v$.}. Furthermore, $P_i$ denotes the $i$th landmark (with $i\in\left\{0,\ldots,67\right\}$ corresponding to facial landmarks and $i\in\left\{68,69\right\}$ the shoulders); $\overrightarrow{PQ}$ denotes the vector between two points $P$ and $Q$, $P^\mu_{i,j}$ is the midpoint between $P_i$ and $P_j$; ${\perp}u$ is the perpendicular vector to $u$ (say, taken clockwise); $\left\Vert\cdot\right\Vert$ is the $L^2$ norm; and $\left\Vert P,u\right\Vert$ is the  $L^2$ distance between a point $P$ and the line spanned by a vector $u$. (Distances can are computed in pixels, but the final geometric measures based on distance are unitless because they are normalized, as described next.)

\subsection{Definitional choices}
\vspace{-1mm}
The descriptions in \tblref{geometric-quantities} are lightly adapted from the source texts for clarity and uniformity. These descriptions were then formalized into the geometric definitions in \tblref{geometric-quantities} by assigning face and body features to landmarks used in our models (as in \figref{face-numbers}), with choices in specific cases as follows:
\begin{itemize}
    \item For facial angle (fa) and habitual head deviation (hhd), the eye line is interpreted as being the line between the midpoints of the eyes, as defined for each eye by the midpoint of its corners, to ensure consistency regardless of whether the eye is open or closed. 
    \item For relative face size (rfs), \cite{akbari_facial_2015} defines this to be the greater divided by the lesser of the two outer-cathnus-to-mouth-corner lengths, but we choose to remove the extra ``logical'' qualifier and always divide the left length by the right length to enable more precise evaluation of predictions. 
    \item For gaze angle (ga) and translational deformity (td), we take the perpendicular bisector to the two shoulder joints as the midsternal plumb line.
    \item For translational deformity (td), since we do not have the luxury of standardized photograph sizes as in \cite{bhaskar_congenital_2017}, we normalize by the distance between the outer cathnuses (eye corners), a relatively stable quantity.
\end{itemize}

\subsection{Omissions}
\vspace{-1mm}
We do not include the following two measurements found in our literature search. The ``facial bulk mass'' of one side compared to the other from \cite{HELVESTON19921609} is alluded to in \cite{akbari_facial_2015} but not precisely defined, and is perhaps not intended to be inferred from photographs. The ``nasal tip deviation'' from \cite{wilson_hoxie_1993} and described in \cite{akbari_facial_2015} is also not made fully precise, and was deemed too difficult to model reliably. Note that we do \textit{not} exclude quantities simply because they cannot be accurately measured algorithmically  with our methods (and indeed, we find that the orbit slopes angle (osa) cannot be).

\section{EXPERIMENTAL SETUP: INFANT DATA AND POSE ESTIMATION}
\vspace{-1mm}

\subsection{Selecting and annotating infant faces}
\vspace{-1mm}

As the testbed for our study, we work with the infant annotated faces (InfAnFace) dataset \cite{wan_infanface_2022}, a comprehensive dataset of 410 infant faces labeled with 68 facial landmark coordinates and four binary pose attributes, designed specifically to alleviate the shortage of annotated data in the infant face domain. The images from the InfAnFace are captured from internet image and video sources, and represent a diverse range of infants in natural environments ``in the wild.'' We manually select a subset of images from InfAnFace Test satisfying the requirements from \secref{assumptions} that the infant be fully front facing, leaving us with 36 images to work with. We manually augment the ground truth facial landmark labels with two additional labels for shoulder joints, to enable us to apply the definitions in \tblref{geometric-quantities}.

\subsection{Face and body landmark estimation}
\vspace{-1mm}

\tblestimators

To study the effectiveness of algorithmic assessment of our geometric measures, we perform experiments with two sets of pose estimation models, the first established models trained largely on adult data, and the second recent models designed specifically for the infant domain. The models and training data are summarized in \tblref{estimators}. 

 In the adult domain, for facial landmark estimation, we employ the high-resolution network (HRNet) \cite{SunXLW19, WangSCJDZLMTWLX19}, an influential multi-resolution convolutional neural network designed to maintain high-resolution representations throughout inference, which is handy for landmark estimation for high-resolution images. For body joint estimation, we use DarkPose \cite{Zhang_2020_CVPR}, which modifies the standard convolutional heatmap regression approach to landmark estimation to alleviate bias in coordinate representation arising from arbitrary pixel quantization during encoding and decoding. Specifically, we adopt a DarkPose model with an HRNet backbone.

In the infant domain, for facial landmark estimation, we use the HRNet-R90JT model from \cite{wan_infanface_2022}, also built on HRNet and adapted to the infant domain via fine-tuning and data augmentation directed at the unique features of infant faces. Finally, for body joint estimation, we use the fine-tuned domain-adapted infant pose (FiDIP) model from \cite{huang_invariant_2021}, which uses synthetic data and domain adversarial methods to adapt body landmark estimation (in this case, the HRNet-backed DarkPose model) to the infant domain.

\section{RESULTS: PERFORMANCE AND ANALYSIS}
\label{sec:results}
\vspace{-1mm}


\subsection{Performance metrics}
\vspace{-1mm}

Often in machine learning, metrics are chosen to enable robust comparison of ever-improving models over time, on large datasets serving as public benchmarks. In our case, because of the relative novelty of the techniques, estimation task, and test set, and because of our interest in immediate medical applications, we rely on a range of more interpretable metrics to give us a sense of absolute performance without needing comparisons. In particular, we will evaluate performance using Spearman's $\rho$ correlation coefficient, the binary classification accuracy of predicting whether a quantity is above or below the mean, and the mean absolute error. We do also report the root mean squared error, for purposes of robust future comparison. 

Spearman's $\rho$ rank correlation coefficient between two variables is defined as the Pearson's $r$ correlation between the internal rankings for each variable. In the context of deep learning tools trained and tested on imprecise labels and small data domains, high Pearson's $r$ correlations between predictions and ground truth is too much to ask for, but adopting Spearman's $\rho$ allows us to relax the requirement for pinpoint accuracy and focus on the relative ranked correctness of the predictions, which is fully compatible with our goal of enabling screening or diagnosis of torticollis. In a similar vein, we assess the \textit{binary classification accuracy} (BCA) of a set of predictions $\left(\tilde{y}_i\right)_{i\in I}$ relative to the ground truth $\left(y_i\right)_{i\in I}$ of a given variable, defined simply as the proportion of $i$ in the sample index $I$ for which $\tilde{y}_i > \mu \iff y_i > \mu$, where $\mu={\mu}\left(\left(y_i\right)_{i\in I}\right)$ is the ground truth sample mean. This coarse measure gauges how often predictions accurately determine whether a quantity is greater or less than the ground truth mean (so unlike Spearman's $\rho$, it is not purely relative), and its simplicity allows for ease of interpretation. We also measure the mean absolute error (MAE) and the root mean squared error (RMSE)---the former being more interpretable, and the latter being more robust for subsequent comparisons. 

\subsection{Results and discussion}
\vspace{-1mm}

\tblperformance 

\tblref{performance} tabulates estimation performance of our geometric measures of symmetry, as gauged by the aforementioned metrics, for both the infant- and adult-based models. 
See \tblref{spearman} for a guideline for interpreting Spearman's $\rho$ scores, from the often cited \cite{dancey2007statistics}.

\tblspearman

On the whole, our results show that the infant pose estimation models yield predictions of five of the six measures---all except the orbit slopes angle (osa)---with a high degree of fidelity. Spearman's $\rho$ correlations are generally strong or very strong and binary classification accuracies high (75.0--88.9\%). Mean absolute errors are within $3^\circ$ for angles and 0.1 for ratios, and root mean squared errors are low as well. The predictions of the remaining quantity, the orbit slopes angle (osa), only moderately correlate with the ground truth, with $\rho=0.36$. For the orbit slopes angle, the BCA is no better than random guessing at 50\%, but the low MAE and RMSE values suggest the poor classification accuracy may be due to miscalibration. Geometrically, the orbit slopes angle is quite subtle, measuring slight differences between the line connecting the outer cathnuses (eye corners) and the inner cathnuses, and it appears that this distinction is too fine to be reliably measured with the infant landmark estimation model. 

Turning to the adult pose estimation models, \tblref{performance} shows that \textit{for every geometric symmetry measure and every prediction performance metric, the predictions from the adult models fare worse than those from the infant models}, with the exception of one tie. Note that this superiority is not maintained when cross-comparing performances of \textit{different} measures---for instance, the infant model prediction of face angle (fa) has lower Spearman's $\rho$ and BCA scores than the adult prediction of gaze angle (ga). Indeed, the adult model predictions of half of the measurements---the relative face size (rfs), gaze angle (ga), and habitual head deviation (hhd)---can be considered to be successful under Spearman's $\rho$ and BCA, although their absolute errors (MAE and RMSE) are still high.

\figscatter

\figpredictions

To complement the summary statistics from \tblref{performance}, we include full scatter plots of predicted vs ground truth values for all six measures under both sets of models, in \figref{scatter}. The scale of these scatter plots is chosen to highlight the presence of major outlier mis-predictions from the adult models, and the subtler performance differences between measurements revealed by \tblref{performance} are harder to perceive. Finally, a visualization of ground truth and predictions the geometric elements involved in the computation of our six measures of interest can be found \figref{predictions}. The predictions from the infant model appear to be fairly accurate, but since it is hard to visually determine angles and ratios at a glance, it is difficult to make definitive conclusions about performance on this qualitative basis alone. What such figures and the performance scatter plots do make it clear, though, is that a major cause of poor performance from the adult pose estimation models stems from a handful of instances of failed facial landmark estimation.

\section{CONCLUSION}
\vspace{-1mm}

Motivated by applications to the detection and treatment of torticollis in infants, we have studied the extent to which recent deep learning based computer vision algorithms designed specifically for the infant domain can measure a set of geometric facial and upper-body symmetry measures previously defined by torticollis researchers, from still images alone. After carefully crafting a test set of infant faces and honing the definitions to line up with standard facial landmarks, we found that most of these measurements can be successfully predicted by pose estimation models designed for infants. We also showed that these models outperform corresponding models designed for and trained on adult data, demonstrating the importance of using tailored neural network models in the data-scarce infant domain. 

One possible next step for computer vision research into torticollis would be to use these predicted geometric measurements as features in machine learning algorithms for screening or diagnosis of torticollis, on clinical subjects or subjects ``in the wild,'' supervised by assessments from medical experts. The underlying data could take the form of either still images or videos. Another potential direction would be to develop these estimation techniques for infants in wider range of poses, not restricted to those clearly facing forward. This would likely require advances in infant pose estimation, especially three-dimensional infant facial landmark estimation, which is currently out of reach.

{\tiny
\bibliographystyle{ieee}
\bibliography{ref}
}

\end{document}